\documentclass[aps,prl,amsmath,amssymb,reprint,groupedaddress]{revtex4-2}

\usepackage{subcaption}
\usepackage{graphicx}
\usepackage{amsmath,amsfonts,amsthm,amssymb,bm,mathtools}
\usepackage{physics}
\usepackage{siunitx}
\usepackage{tikz}
\usepackage{definitions}
\usepackage[LGR,T1]{fontenc}
\usepackage[utf8]{inputenc}
\usepackage{hyperref}
\hypersetup{
  colorlinks=true,
  urlcolor=blue
}
\usepackage{cleveref}

\DeclareRobustCommand{\tikzfilledcircle}{%
  \tikz[baseline=-0.6ex]{\fill (0,0) circle (0.5ex);}%
}
\DeclareRobustCommand{\tikzcircle}{%
  \tikz[baseline=-0.6ex]{\draw (0,0) circle (0.5ex);}%
}
\DeclareRobustCommand{\tikzfilledtriangle}{%
  \tikz[baseline=-0.6ex]{\fill (0,-0.5ex) -- (1ex,-0.5ex) -- (0.5ex,0.5ex) -- cycle;}%
}
\DeclareRobustCommand{\tikzfilledsquare}{%
  \tikz[baseline=-0.6ex]{\fill (-0.5ex,-0.5ex) -- (0.5ex,-0.5ex) -- (0.5ex,0.5ex) -- (-0.5ex,0.5ex) -- cycle;}%
}

\newcommand{\hollowlinesymbol}{%
\tikz[baseline=-0.6ex]{
  \draw[thick] (0,0) -- (0.6,0);
  \draw[thick, fill=white] (0.3,0) circle (1.5pt);
}}

\begin{document}
\preprint{}

%Title of paper
\title{Elastic pseudoturbulence induced by low-Galilei settling spheres}

\author{Ludovico Foss\`{a}\textsuperscript{1}}
\email[Contact author ]{ludovico-fossa@oist.jp}
\author{Michele Macaluso\textsuperscript{1,2}}
\author{Luca Brandt\textsuperscript{2}}
\author{Marco Edoardo Rosti\textsuperscript{1}}
\email[Contact author ]{marco-rosti@oist.jp}

\affiliation{\textsuperscript{1}Complex Fluids and Flows Unit, Okinawa Institute of Science and Technology Graduate University, 1919-1 Tancha, Onna-son, Kunigami-gun, Okinawa-ken, Japan \\ \textsuperscript{2}Dipartimento di Ingegneria dell'Ambiente, del Territorio e delle Infrastrutture (DIATI), Politecnico di Torino, Corso Duca degli Abruzzi 26, 10129 Torino (TO), Italy}

\date{\today}

\begin{abstract}
In this Letter, we show how a suspension of light solid spheres settling through a polymer solution results in a chaotic and highly-intermittent state. By leveraging particle-resolved direct numerical simulations, we investigate the effect of increasing polymer relaxation time and Deborah number $\Deborah$ on viscoelastic sedimentation at a low density ratio $\rho_s/\rho_f=5$ and a low Galilei number $\Galilei=3.16$. Even at moderate $\Deborah$, the spheres form gravity-aligned clusters and settle faster, while the polymer stresses energize the large scales of motion. The onset of elastic turbulence and intermittency is signaled by a $-4$ spectral scaling in the high-wavenumber range and by nonlinear high-order exponents of the velocity structure functions. These results indicate that viscoelastic effects induce pseudoturbulence in the presence of viscous-dominated sedimentation.
\end{abstract}

\maketitle

\textit{Introduction}. Sedimentation in liquids and gases lies at the core of several industrial applications and natural phenomena \citep{Guazzelli_Hinch_2011,Guazzelli_Morris_Pic_2011,Marchioli_Bourgoin_Coletti_Fox_Magnaudet_Reeks_Simonin_Sommerfeld_Toschi_Wang_Balachandar_2025,Daugan_Talini_Herzhaft_Peysson_Allain_2004,Jeong_Urgeles_Bahk_Yoo_Lee_2022,Whorton_Jones_Russell_2025}. As they settle, solid particles shed wakes, transferring momentum and energy to an otherwise quiescent fluid that may transition to a chaotic state. This state, commonly referred to as pseudoturbulence \citep{Brandt_Coletti_2022}, bears similarities to that induced by rising bubbles \citep{Theofanous_Sullivan_1982,Lance_Bataille_1991} and is characterized by flow structures much larger than the particle size \citep{Risso_2018,Ravisankar_Zenit_2025,Jiang_Brandt_Xu_Zhao_2025}. For sphere-laden incompressible Newtonian fluids, the onset of pseudoturbulence, the formation of clusters and the enhancement of settling is governed by the solid-to-fluid density ratio $\rho_s/\rho_f$, the suspension's volume fraction $\Phi_s$ and the relative strength of gravity over viscosity, which is quantified by the Galilei number $\Galilei$ \citep{Uhlmann_Doychev_2014,Fornari_Wade_Brandt_Picano_2019,Jiang_Mirzareza_Crialesi-Esposito_Brandt_2026}. 
The physics of suspensions are even more intricate when fluids with complex rheological properties, such as polymer solutions, are considered \citep{DAvino_Maffettone_2015,Zenit_Feng_2018}. In a viscoelastic fluid, polymer chains stretch under shear, storing elastic energy that is released into the solvent as they relax. This backreaction is governed by the polymer degree of dilution and relaxation time: when the ratio of the latter to the fluid time scale (quantified by the Deborah number $\Deborah$ \citep{Zenit_Feng_2018}) is sufficiently high, the flow becomes chaotic even in the limit of negligible inertia \citep{Groisman_Steinberg_2000,Steinberg_2021,Garg_Rosti_2025}. This state, referred to as elastic turbulence, is characterised by a balance between the work of the polymeric stresses and viscous dissipation and by an energy spectrum $E(\kappa)\sim\kappa^{-4}$ \citep{Singh_Perlekar_Mitra_Rosti_2024,Singh_Rosti_2025}. 

Laboratory experiments involving a large number of spheres have shown the particle's tendency to cluster in gravity-aligned columns at small $\Galilei$ and small $\rho_s/\rho_f$ \citep{Allen_Uhlherr_1989,Bobroff_Phillips_1998,Mora_Talini_Allain_2005}. These streamers, which result in enhanced settling, are not observed under the same conditions in Newtonian fluids \citep{Guazzelli_Hinch_2011,Brandt_Coletti_2022}. Hence, this letter aims to answer the following question: can polymer elasticity feed energy to scales larger than the particles and induce pseudoturbulence in suspensions of quasi-neutrally buoyant, low-$\Galilei$ spheres? Indeed, the recent experiments by \citet{Ravisankar_Zenit_2025} have shown that air bubbles dispersed in aqueous polymer solutions at large $\Galilei=O(10^3)$ modulate turbulence at scales larger than their diameter. Prompted by these results, we investigate the sedimentation of solid particles at density ratios and $\Galilei$ of order unity, a regime in which a Newtonian fluid exhibits smooth settling. We demonstrate, using state-of-the-art particle-resolved direct numerical simulations (PR-DNSs), that increasing polymer elasticity results in the development of a chaotic and intermittent three-dimensional flow. 

\begin{figure*}
    \includegraphics[width=\linewidth]{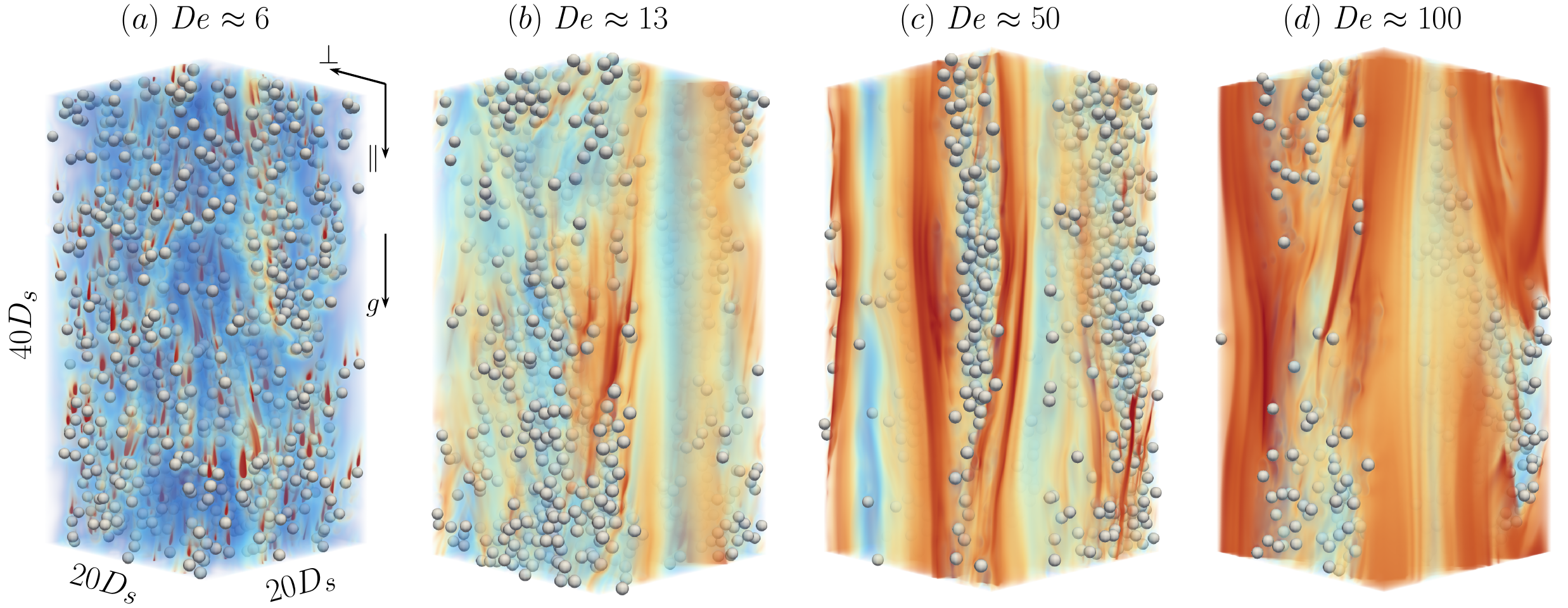}
    \caption{The domain is a parallelepiped $20D_s\times20D_s\times40D_s$ and gravity is oriented along the longest edge. The symbols $\parallel$ and $\perp$ denote the vertical direction (parallel to gravity) and horizontal direction (perpendicular to gravity), respectively. Each panel shows a volumetric rendering of the trace of the conformation tensor $C_{kk}$ for $\Deborah\approx6$ (\textit{a}), $13$ (\textit{b}), $50$ (\textit{c}) and $100$ (\textit{d}). Regions of low and high $C_{kk}$ are drawn in shades of blue and red, respectively.}
    \label{fig:trace_conformation_tensor}
\end{figure*}

\textit{Methodology}. We consider a swarm of spheres settling through the triply periodic fluid domain shown in figure \ref{fig:trace_conformation_tensor}, with gravity as the only external forcing. The latter consists of a parallelepiped with sides of length $20 \times 20 \times 40$ in units of the sphere diameter $D_s$.
The acceleration due to gravity $g$ is directed along the longest edge and the spheres occupy $\Phi_s=5\%$ of the total volume \citep{Murch_Shaqfeh_2020}. The fluid velocity field $u_i$ is solenoidal and governed by the Navier--Stokes equations, with the peculiarity that the stress tensor also distills the contribution of the polymeric stresses $\mu_{\polymer} ( C_{ij} - \delta_{ij})/\tau_{\polymer}$ in addition to that of the fluid pressure $p$ and the Newtonian stresses $\mu_f(\p_ju_i+\p_iu_j)$ \citep{DAvino_Maffettone_2015,Izbassarov_Rosti_Ardekani_Sarabian_Hormozi_Brandt_Tammisola_2018}. Here, $\mu_f$ and $\mu_\polymer$ are the fluid and polymer dynamic viscosity, respectively, and $\tau_{\polymer}$ is the polymer relaxation time \citep{FujinNote}. The conformation tensor $C_{ij}$ is symmetric and governed by the Oldroyd-B model \cite{Oldroyd_1950,Leslie_Tanner_1961,Feng_Huang_Joseph_1996,DAvino_Maffettone_2015,Izbassarov_Rosti_Ardekani_Sarabian_Hormozi_Brandt_Tammisola_2018}. Although this model admits infinite extensibility, the present results show a peak elongation $O(10^2)$ comparable to the values measured in laboratory experiments \cite{[See the values of $\Lmax$ in table Table S2 and Figure S5 in the Supplemental Material of ]Yamani_Keshavarz_Raj_Zaki_McKinley_Bischofberger_2021}. The temporal evolution of the spheres' translational velocity $U_i$ is governed by the Newton-Euler equations \citep{Hori_Takagi_Rosti_2022}. The system is governed by four independent parameters: the solid-to-fluid density ratio $\rho_s/\rho_f$, the viscosity ratio $\beta=\mu_f\left(\mu_f+\mu_\polymer\right)^{-1}$, the Galilei number $\Galilei\equiv  \rho_fU_sD_s/(\mu_f+\mu_\polymer)$ and the Deborah number $\Deborah\equiv \tau_\polymer U_s / D_s$, where $U_s\equiv\vert(\rho_s/\rho_f-1)g\vert^{1/2}D_s^{1/2}$ is the sedimentation velocity \cite{Uhlmann_Doychev_2014,DAvino_Maffettone_2015,Brandt_Coletti_2022} and $D_s/U_s$ is the gravitational time \citep{Uhlmann_Doychev_2014}. To examine the effect of polymer elasticity in a low-inertia regime, we consider constant $\rho_s/\rho_f=5$ \citep{Allen_Uhlherr_1989,Murch_Shaqfeh_2020}, $\beta=0.5$ and $\Galilei=3.16$ while varying $\tau_\polymer$ to increase $\Deborah$. Five cases with $\Deborah\approx 1$, $6$, $13$, $50$, and $100$ are compared with the benchmark Newtonian case ($\Deborah=0$, $\beta=1$). All simulations were performed using the in-house parallel Fortran code \href{https://www.oist.jp/research/research-units/cffu/fujin}{\textit{Fujin}} \citep{Hori_Takagi_Rosti_2022,Rosti_2026,FujinNote}. 

\begin{figure}
    \includegraphics[width=\linewidth]{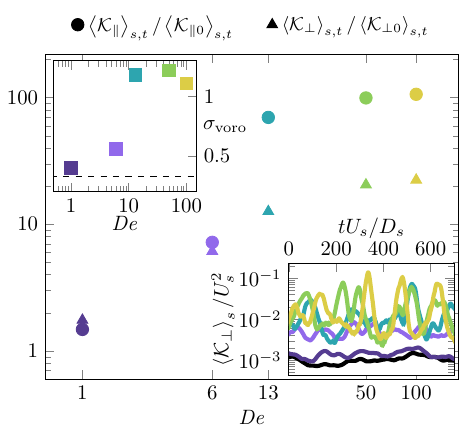}
    \caption{Ensemble-time averages $\averagespheretime{\cdot}$ of the  horizontal, $\mathcal{K}_\perp$, (\tikzfilledtriangle) and vertical, $\mathcal{K}_\parallel$ (\tikzfilledcircle), kinetic energy of the spheres, normalised on the Newtonian case, for increasing $\Deborah$. Bottom-right inset: time histories of the volume-averaged horizontal kinetic energy $\averagesphere{\mathcal{K}_\perp}$. Top-right inset: standard deviation of the Voronoi volumes $\sigmavoro$ (\tikzfilledsquare) along with the values corresponding to a Newtonian fluid and to a random distribution of particles $\approx0.33$ (dashed line).}
    \label{fig:particles}
\end{figure}

\textit{Results}. As shown in figure \ref{fig:trace_conformation_tensor}, increasing the relaxation time $\tau_\polymer$ alters the dynamics of the fluid phase and the motion of the spheres significantly. The visualizations display four instantaneous renderings of the trace of the conformation tensor $C_{kk}$ and the spheres for $\Deborah\approx6$ (\textit{a}), $13$ (\textit{b}), $50$ (\textit{c}) and $100$ (\textit{d}). As $\Deborah$ increases, the polymers stretch more and more while the spheres cluster in gravity-oriented columns \citep{Allen_Uhlherr_1989,Bobroff_Phillips_1998,Mora_Talini_Allain_2005}. At low $\Deborah$ (\textit{a}), the high-$C_{kk}$ regions are confined to the viscoelastic wakes of the spheres, while at high $\Deborah\gtrsim13$ (\textit{b}, \textit{c}, \textit{d}) these wakes are stretched along the gravity direction and fill the regions of low sphere concentration. \cite{Monchaux_Bourgoin_Cartellier_2010,Uhlmann_Doychev_2014,VoronoiNote} To quantify the spheres' tendency to concentrate \citep{Mora_Talini_Allain_2005}, we compute the Voronoi tessellation of the suspension. The degree of clustering (i.e. the normalised standard deviation of the Voronoi volumes $\sigmavoro$) is plotted in the top-left inset in figure \ref{fig:particles} and compared with that obtained for a random distribution of spheres. The standard deviation is the same for the random and Newtonian cases ($\sigmavoro\approx0.33$), but increases markedly with $\Deborah$, indicating the formation of clusters \cite{VoronoiNote}.

Increasing polymer elasticity also enhances the settling and horizontal motion of the particles. The ensemble-time averages $\averagespheretime{\cdot}$ associated to the horizontal velocity components ($\mathcal{K}_\perp$, triangles) and vertical velocity components ($\mathcal{K}_\parallel$, circles) of a sphere's kinetic energy $\mathcal{K}=U_iU_i/2$ are normalised on their Newtonian values $\averagespheretime{\mathcal{K}_{\perp0}}$ and $\averagespheretime{\mathcal{K}_{\parallel0}}$ and plotted in figure \ref{fig:particles} as a function of $\Deborah$. For all $\Deborah$, the kinetic energy increases significantly with respect to the corresponding Newtonian values. For large $\Deborah\gtrsim13$, it saturates at a 20-fold enhancement of the horizontal kinetic energy and a 100-fold enhancement of the vertical component. Interestingly, the time history of the ensemble average $\averagesphere{\mathcal{K}_\perp}$ (bottom-right inset in figure \ref{fig:particles}) shows strong localized bursts whose amplitude increases with $\Deborah$. 

\begin{figure}
    \includegraphics[width=\linewidth]{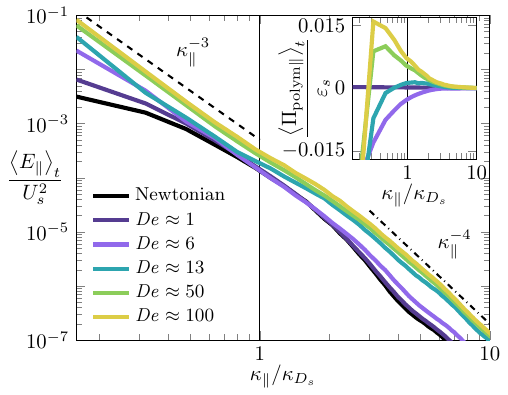}
    \caption{Normalised vertical-velocity energy spectra ${E_{\parallel}}/U_s^2$ for increasing $\Deborah$. The vertical line denotes the spheres' wavenumber $\kappa_{D_s} = 2\pi/D_s$, while the dashed and dash dot lines describe the $-3$ and $-4$ scaling, respectively. The inset shows the polymer term $\left\langle\Pi_{\polymer\parallel}\right\rangle$ in the scale-by-scale kinetic energy budget. The latter is normalized by a factor $\varepsilon_s = \mu_f(U_s/D_s)^2$.}
    \label{fig:spectra}
\end{figure}

Thus far, we have shown that increasing $\tau_\polymer$ amplifies the elastic energy absorbed by the polymers (see $C_{kk}$ \citep{Rosti_Perlekar_Mitra_2023}) and enhances settling. To understand how elasticity modulates the fluid flow, we proceed by examining the kinetic energy spectra. Because of the flow anisotropy, we focus on the spectra associated to the velocity fluctuations along gravity $E_\parallel\left(\kappa_\parallel\right)$ shown in figure \ref{fig:spectra} \citep{SpectraNote}. Viscoelasticity enhances the energy content across the spectrum and leads to the onset of two characteristic scalings which are not observed in the Newtonian case (black line). At sub-particle scales ($\kappa_\parallel/\kappa_{D_s}>1$), the spectra display a $-4$ slope typical of elastic turbulence \cite{Steinberg_2021,Singh_Perlekar_Mitra_Rosti_2024,Garg_Rosti_2025,Singh_Rosti_2025}, while at the large scales ($\kappa_\parallel/\kappa_{D_s}<1$), they follow a $-3$ slope, similarly to what has been widely reported in multiphase dispersed flows \cite{Riboux_Risso_Legendre_2010,Roghair_Martinez-Mercado_van-Sint-Annaland_Kuipers_Sun_Lohse_2011,Risso_2018,Pandey_Mitra_Perlekar_2023,Ravisankar_Zenit_2025,Jiang_Mirzareza_Crialesi-Esposito_Brandt_2026}. The spectra of the transverse kinetic energy $E_\perp\left(\kappa_\perp\right)$ display the same low and large-wavenumber scalings, see supplementary material \cite{SpectraNote}.

Strikingly, increasing elasticity energizes the fluid motion at scales larger than the particle diameter ($\kappa_\parallel/D_s<1$). To understand this modulation, we consider the scale-by-scale energy balance for the fluid phase \citep{Pope_2000}, which is obtained by contracting the Fourier transform of the fluid momentum balance with the transformed velocity $\hat u_i$ and averaging over $\kappa_{\perp}$ \cite{Pandey_Mitra_Perlekar_2023,Singh_Rosti_2025,Ravisankar_Zenit_2025,Jiang_Mirzareza_Crialesi-Esposito_Brandt_2026,SpectraNote}. The contribution of the polymer stresses in the gravity direction is isolated in the term $\Pi_{\polymer\parallel}$ \cite{SpectraNote}, whose temporal average $\averagetime{\Pi_{\polymer\parallel}}$ is displayed in the inset of figure \ref{fig:spectra}. Its magnitude is negligible for $\Deborah\leq1$, consistent with the weak departure of $E_\parallel$ from the Newtonian case (figure \ref{fig:spectra}), negative for $\Deborah\lesssim13$ and positive for larger $\Deborah$. In this limit, $\Pi_{\polymer\parallel}$ increases markedly for $\kappa_\parallel/\kappa_{D_s}<1$, indicating that the polymer stresses energize the scales larger than the sphere diameter as $\tau_\polymer$ increases. Albeit differing in magnitude, the work of the polymer stresses in the horizontal  directions, $\Pi_{\polymer\perp}$, shows the same monotonic growth with $\Deborah$ \cite{SpectraNote}. Thus, the polymer stresses mediate the transfer of energy between the settling spheres and the large scales of the fluid motion, with the polymers injecting energy at scales larger than $D_s$.

\begin{figure}
    \includegraphics[width=\linewidth]{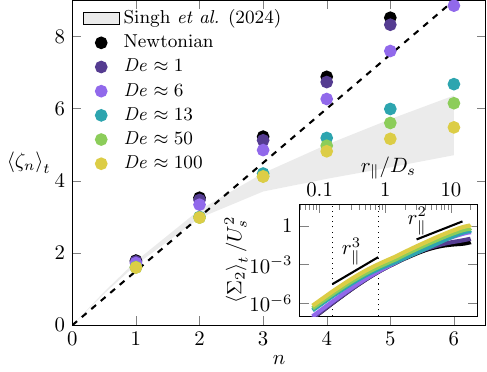}
    \caption{Exponents $\zeta_n$ of the of the structure functions of order $n$ \citep{Biferale_Cencini_Lanotte_Vergni_Vulpiani_2001,Singh_Perlekar_Mitra_Rosti_2024} at small separations $r_\parallel$ (\tikzfilledcircle). The results are compared with those of \citet{Singh_Perlekar_Mitra_Rosti_2024} (gray area). Inset: second-order structure function $\Sigma_2$ and the scaling laws $\zeta_2=3$ and $\zeta_2=2$ at small and large separations, respectively. The range used to extract $\zeta_n$ is delimited by the dotted lines.}
    \label{fig:exponents}
\end{figure}

The results shown in figure \ref{fig:spectra} prompt two questions. The first is whether the observed scalings are really indicative of elastic turbulence; the second is whether the increase in fluid kinetic energy and the intensification of polymer stresses at large scales are related the length of the sphere's viscoelastic wakes observed in figure \ref{fig:trace_conformation_tensor}. As for the first question, a $-4$ scaling has also been reported in Newtonian laminar and turbulent flows laden with small spheres \citep{Wang_Ayala_Gao_Andersen_Mathews_2014,Ramirez_Burlot_Zamansky_Bois_Risso_2024,Chiarini_Tandurella_Rosti_2025,Jiang_Mirzareza_Crialesi-Esposito_Brandt_2026}. To verify the nature of the observed $\kappa_\parallel^{-\alpha}$ power-law decay (with $\alpha=4$) at high $\kappa_\parallel$ \cite{Rosti_Perlekar_Mitra_2023,Singh_Rosti_2025}, we compute the velocity structure functions and extract the intermittency exponents for comparison with those of elastic turbulence. Because the $\alpha=4$ scaling lies in the range of one-time differentiability \cite[\S G]{Pope_2000}, the velocity field is analytic for small separations $r_\parallel$ and the longitudinal structure function $S_n(r_\parallel,t)=\averagespace{\left[\Delta_\parallel\left(r_\parallel,t\right)\right]^n}$ (where $\Delta_\parallel(r_\parallel,t) = u_\parallel(\boldsymbol{x} + r_\parallel\boldsymbol{e}_\parallel,t) - u_\parallel(\boldsymbol{x},t)$ and $\averagespace{\cdot}$ denotes spatial averaging) expands at leading-order as $S_n\sim r_\parallel^n$ for small $r_\parallel$, regardless of intermittency \cite{Biferale_Cencini_Lanotte_Vergni_Vulpiani_2001,Schumacher_Sreenivasan_Yakhot_2007,Garg_Rosti_2025}. If present, intermittency is expected to affect the higher-order terms of $S_n$. We thus employ the method used by \citet{Singh_Perlekar_Mitra_Rosti_2024} to extract the scaling exponents $\zeta_n$ from the structure functions of the second velocity differences $\Sigma_n(r_\parallel,t)=\averagespace{[\Delta_\parallel^2(r_\parallel,t)]^n} \sim r_\parallel^{\zeta_n}$ (where $\Delta_\parallel^2(r_\parallel,t) = u_\parallel(\boldsymbol{x} + r_\parallel\boldsymbol{e}_\parallel,t) - 2u_\parallel(\boldsymbol{x},t) + u_\parallel(\boldsymbol{x}-r_\parallel\boldsymbol{e}_\parallel,t)$ \citep{Biferale_Cencini_Lanotte_Vergni_2003,Singh_Perlekar_Mitra_Rosti_2024,Garg_Rosti_2025}) for which theory predicts $\zeta_n=n\left(\alpha-1\right)/2$ in the absence of intermittency \cite{Bernard_2000,Biferale_Cencini_Lanotte_Vergni_Vulpiani_2001,StructureFunctionNotes}. The computed exponents $\zeta_n$ are shown in figure \ref{fig:exponents} for $n=1$ to $6$. They vary linearly with $n$ in the Newtonian and low-$\Deborah$ cases, confirming the absence of intermittency. For $\Deborah\gtrsim13$, they deviate significantly from the theoretical prediction $3n/2$ (dashed line) and vary nonlinearly with $n$ at large $n$, indicating the onset of intermittency \citep{Pope_2000,Schumacher_Sreenivasan_Yakhot_2007,Iyer_Sreenivasan_Yeung_2020}. At $\Deborah \gtrsim 50$ (green and yellow circles), the exponents lie within the range reported by \citet{Singh_Perlekar_Mitra_Rosti_2024} (gray shaded area) for  elastic, homogeneous and isotropic turbulence, thus proving the presence of elastic turbulence at the small scales of our suspension flow. Note that, as shown in the inset of figure \ref{fig:exponents}, the second-order  structure functions of the second differences $\Sigma_2(r_\parallel)$ indeed scale as $r_\parallel^3$ and $r_\parallel^2$ for small and large separations, respectively, consistent with the spectral scalings $-4$ and $-3$ shown in figure \ref{fig:spectra} \citep{StructureFunctionNotes}. The turbulent spectra and the structure functions indicate the onset of a multiscale, chaotic and highly intermittent particle-laden viscoelastic flow, with two distinct turbulent states at different scales.

\begin{figure}
    \includegraphics[width=\linewidth]{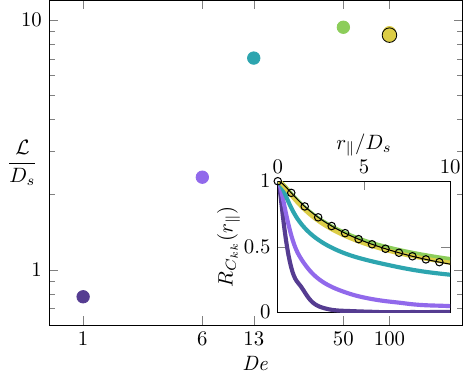}
    \caption{Length of the viscoelastic wakes $\mathcal{L}$ for increasing $\Deborah$ (\tikzfilledcircle), together with the results obtained in a taller domain of size $20D_s\times20D_s\times80D_s$ (\tikzcircle). Bottom-right inset: parallel autocorrelation $R_{C_{kk}}(r_\parallel)$ as a function of $r_\parallel$ (solid curves), as well as that obtained in a domain $20D_s\times20D_s\times80D_s$ (\protect\hollowlinesymbol).}
    \label{fig:autocorrelations}
\end{figure}

Turning to the second question -- the link between enhanced elastic energy and the lifetime of the viscoelastic wakes -- if this hypothesis is correct, the spatial correlation of regions with high $C_{kk}$ should increase along the vertical direction as $\Deborah$ increases. We therefore compute the integral length $\mathcal{L}=\int_0^\infty \left[R_{C_{kk}}\left(r_\parallel\right)/R_{C_{kk}}\left(0\right)\right] \differential r$ based on the vertical autocorrelation $R_{C_{kk}}\left(r_\parallel\right)=\averagespacetime{C_{kk}\left(\boldsymbol{x},t\right)C_{kk}\left(\boldsymbol{x}+r_\parallel\boldsymbol{e}_\parallel,t\right)} - \averagespacetime{C_{kk}\left(\boldsymbol{x},t\right)}^2$ (where the operator $\averagespacetime{\cdot}$ denotes average in space and time) and plot the results in figure \ref{fig:autocorrelations}. Indeed, the integral length $\mathcal{L}$ increases monotonically up to $\Deborah\approx50$ (circles in figure \ref{fig:autocorrelations}), and appears to saturate for $\Deborah\approx100$. As shown in the inset in figure \ref{fig:autocorrelations}, the vertical autocorrelation of $C_{kk}$ decays over a few sphere diameters at small $\Deborah$, but persists over larger distances at high $\Deborah$. To verify that the saturation of $\mathcal{L}$ at large $\Deborah$ is independent of the domain finite length, we performed an additional simulation at $\Deborah\approx100$ on a longer domain of size $20D_s\times 20D_s\times 80D_s$. The resulting wake length, depicted by the empty circle in the figure, differs only by 2\% from that computed on the reference domain, see figure \ref{fig:trace_conformation_tensor}. To summarize, increasing $\Deborah$ leads to the formation of gravity-aligned streamers. Separate wakes merge and fill the sphere-void regions adjacent to the streamers. When relaxing, the polymers inject energy at scales comparable to those of the streamers, thereby modulating the spectra and enhancing the velocity fluctuations and small-scale intermittency.

In this Letter, we have demonstrated that the sedimentation of low-Galilei spheres in quiescent viscoelastic fluid and in the absence of flow inertia, induces elastic pseudoturbulence already for moderate values of the Deborah number. Settling spheres cluster in gravity-aligned streamers and stretch the polymers in their wakes. As they relax, the polymers inject elastic energy back into the solvent at scales larger than the spheres' diameter. The exponents of the structure functions of the second velocity differences reveal a marked intermittent character, consistent with previously observed elastic homogeneous and isotropic turbulence \cite{Singh_Perlekar_Mitra_Rosti_2024,Garg_Rosti_2025}. Our results suggest that polymer additives may be employed to enhance turbulent mixing and reaction kinetics in the presence of viscous-dominated sedimentation, thereby offering new perspectives on the role of complex fluids in nature and engineering \citep{Daugan_Talini_Herzhaft_Peysson_Allain_2004,Jeong_Urgeles_Bahk_Yoo_Lee_2022,Whorton_Jones_Russell_2025}.

\begin{acknowledgments}
\textit{Acknowledgments}. The research was supported by the Okinawa Institute of Science and Technology Graduate University (OIST) with subsidy funding to M.E.R. from the Cabinet Office, Government of Japan. M.E.R. also acknowledges funding from the Japan Society for the Promotion of Science (JSPS), grant 24K17210 and 24K00810. M.M. also acknowledges the support of the Polytechnic University of Turin's Outgoing Mobility Scholarship (Tesi su Proposta) and the OIST Research Internship Program. The Authors acknowledge the computational resources provided by the Scientific Computing \& Data Analysis Section of the Core Facilities at OIST and by HPCI under Research Project grants hp250021, hp250035, hp260009, and hp260019. L.F., M.M. and M.E.R. are grateful to Dr. Piyush Garg for his comments and suggestions. 
\end{acknowledgments}

\textit{Data availability}. The data that support the findings of this Letter are not publicly available. The data are available from the Authors upon reasonable request.

\nocite{Hori_Takagi_Rosti_2022,Izbassarov_Rosti_Ardekani_Sarabian_Hormozi_Brandt_Tammisola_2018,Sugiyama_Ii_Takeuchi_Takagi_Matsumoto_2011,Fattal_Kupferman_2004,Tsuji_Kawaguchi_Tanaka_1993,Costa_Boersma_Westerweel_Breugem_2015,Monchaux_Bourgoin_Cartellier_2010,Monchaux_2012,Fiabane_Zimmermann_Volk_Pinton_Bourgoin_2012,Lu_Lazar_Rycroft_2023}
\bibliography{biblio,footnotes}

\end{document}

\preprint{}

\title{Supplemental material for \emph{Elastic pseudoturbulence induced by low-Galilei settling spheres}}
\author{Ludovico Foss\`{a}\textsuperscript{1}}
\email[Contact author ]{ludovico-fossa@oist.jp}
\author{Michele Macaluso\textsuperscript{2}}
\author{Luca Brandt\textsuperscript{2}}
\author{Marco Edoardo Rosti\textsuperscript{1}}
\email[Contact author ]{marco-rosti@oist.jp}

\affiliation{\textsuperscript{1}Complex Fluids and Flows Unit, Okinawa Institute of Science and Technology Graduate University, 1919-1 Tancha, Onna-son, Kunigami-gun, Okinawa-ken, Japan \\ \textsuperscript{2}Dipartimento di Ingegneria dell'Ambiente, del Territorio e delle Infrastrutture (DIATI), Politecnico di Torino, Corso Duca degli Abruzzi 26, 10129 Torino (TO), Italy}

\date{\today}

\begin{abstract}
% insert abstract here
\end{abstract}

\maketitle

\tableofcontents
\section{Direct numerical simulations}

\subsection{Mathematical formulation}
The fluid velocity field $u_i$ is solenoidal and governed by the momentum balance 
\begin{equation}\label{eq:navier-stokes}
    \p_t u_i + u_j\p_ju_i = \frac{1}{\rho_f}\p_j\sigma_{ij} + f_i^{\particle} + \left(\frac{\rho_s}{\rho_f}-1\right)g\delta_{zz},
\end{equation}
where $\rho_f$ and $\rho_s$ are the fluid and solid density, $f_i^{\particle}$ is the volume forcing exerted by a sphere and $\delta_{ij}$ is the Kronecker delta. Summation over repeated indices is implied. The stress tensor $\sigma_{ij}$ distills the contribution of the fluid pressure $p$, the Newtonian stresses and the polymeric stresses
\begin{equation}\label{eq:stress_tensor}
    \sigma_{ij} = - p\delta_{ij} + \mu_f \left(\p_ju_i + \p_iu_j\right) + \frac{\mu_{\polymer}}{\tau_{\polymer}} C_{ij}.
\end{equation}
Here, $\mu_f$ and $\mu_\polymer$ are the fluid and polymer dynamic viscosity, respectively, and $\tau_{\polymer}$ is the polymer relaxation time. The conformation tensor $C_{ij}$ is symmetric and governed by a transport equation based on the Oldroyd-B model \cite{Oldroyd_1950,Leslie_Tanner_1961,Feng_Huang_Joseph_1996,DAvino_Maffettone_2015,Izbassarov_Rosti_Ardekani_Sarabian_Hormozi_Brandt_Tammisola_2018}
\begin{equation}\label{eq:oldroyd-b}
    \p_t C_{ij} + u_k\p_k C_{ij} = C_{ik} \p_k u_j + C_{kj} \p_i u_k - \frac{C_{ij} - \delta_{ij}}{\tau_\polymer}.
\end{equation}
The velocity of a point inside a solid sphere is the sum of the translational velocity and the rotation around the center $U_{i} + \levicivita \Omega_{j} R_k$, where $\levicivita$ the Levi-Civita tensor and $R_k$ the radial distance from the particle center. The temporal evolution of $U_{i}$ and $\Omega_i$ is governed by the Newton-Euler equations \cite{Hori_Takagi_Rosti_2022,Rosti_2026}
\begin{subequations}\label{eq:newton_euler}
\begin{equation}
    m_s\frac{\differential U_i}{\p t} = \oint_{\mathcal{V}_s} \frac{\p \sigma_{ij}}{\p x_j}\differential\mathcal{V} + F_i^{\collision} + \left(\rho_s-\rho_f\right)\mathcal{V}_sg\delta_{zz},
\end{equation}
\begin{equation}
    \mathcal{I}_s \frac{\differential \Omega_i}{\differential t} = \oint_{\mathcal{V}_s} \levicivita R_j \p_l\sigma_{kl} \differential\mathcal{V}.
\end{equation}
\end{subequations}
Here, $F^{\collision}$ is the force exerted by inter-particle collisions and $\mathcal{V}_s=\pi D_s^3/6$, $m_s = \rho_s \mathcal{V}_s$ and $\mathcal{I}_s = m_sD_s^2 / 10$ are the sphere's volume, mass and moment of inertia, respectively. A vertical pressure gradient is imposed that results in a zero-mean settling velocity for the fluid phase $\averagespace{u_\parallel}=0$. 

\subsection{Non-dimensional equations and parameters}
The terms in the momentum balance equation \eqref{eq:navier-stokes} and the transport equation of the conformation tensor \eqref{eq:oldroyd-b} are scaled by the settling velocity of the suspension $U_s\equiv\vert(\rho_s/\rho_f-1)g\vert^{1/2}D_s^{1/2}$ \cite{Brandt_Coletti_2022}, the sphere diameter $D_s$ and the total dynamic viscosity $\mu_f+\mu_\polymer$. We recall that $\rho_s/\rho_f$ is the density ratio and $g$ the acceleration due to gravity. The pressure $p$ scales with $\left(\mu_f+\mu_\polymer\right)U_sD_s^{-1}$ and the conformation tensor $C_{ij}$ is non-dimensional by definition. In this section, we denote non-dimensional variables with a tilde $\tilde\cdot$. The governing equations for the carrier flow read, in non-dimensional form
\begin{subequations}
\begin{multline}\label{eq:momentum_balance}
    \Galilei \left( \p_t\tilde u_i + \tilde u_j\p_j\tilde u_i - \tilde f_i^{\particle} + \delta_{zz}\right) = \\ = - \p_i\tilde p + \beta\p_j^2\tilde u_i + \left(1-\beta\right)\Deborah^{-1} \p_j C_{ij},
\end{multline}
\begin{multline}\label{eq:conformation_transport}
    \p_t C_{ij} + \tilde u_k\p_k C_{ij} - C_{ik}\p_k \tilde u_j - C_{kj} \p_i \tilde u_k =\\
    = \left(\delta_{ij} - C_{ij}\right)\Deborah^{-1}.
\end{multline}
\end{subequations}
Here, three non-dimensional parameters appear, namely the Galilei number $\Galilei\equiv \rho_fU_sD_s\left(\mu_f+\mu_\polymer\right)^{-1}$, the Deborah number $\Deborah\equiv U_s\tau_\polymer D_s^{-1}$ and the viscosity ratio $\beta\equiv\mu_f\left(\mu_f+\mu_\polymer\right)^{-1}$. Similarly, the stress tensor $\sigma_{ij}$ scales with $\left(\mu_f+\mu_\polymer\right)U_sD_s^{-1}$ and the Newton-Euler equations read \cite{DAvino_Maffettone_2015}
\begin{subequations}
\begin{equation}
    \Galilei \frac{\rho_s}{\rho_f} \frac{\pi}{6} \left( \frac{\differential \tilde U_i}{\differential \tilde t} - \tilde F_i^{\collision} - \delta_{zz}\right) = \oint_{\mathcal{V}_s} \frac{\p\tilde\sigma_{ij}}{\p\tilde x_j}\differential\tilde{\mathcal{V}} ,
\end{equation} 
\begin{equation}
    \Galilei \frac{\rho_s}{\rho_f} \frac{\pi}{60} \frac{\differential\tilde\Omega_i}{\differential \tilde t} = \oint_{\mathcal{V}_s}\levicivita \tilde R_j\frac{\p\tilde \sigma_{kl}}{\p \tilde x_l}\differential\tilde{\mathcal{V}}.
\end{equation}
\end{subequations}
The dynamics of this physical system are thus determined by four independent parameters: $\rho_s/\rho_f$, $\Galilei$, $\Deborah$ and $\beta$.

\subsection{Simulation parameters}

The computational domain consists of two adjacent cubes of side length $20D_s$ and total volume $\vdomain=2(20D_s)^3=16000D_s^3$. We fixed the gravity $g=-1$, the dynamic viscosity of the solvent $\mu_f=0.1$, the polymer viscosity ratio $\beta=0.5$, the number of spheres $N_s=1528$, the sphere diameter $D_s=1$ and the solid-to-fluid density ratio $\rho_s/\rho_f=5$. The volume fraction of the dispersed phase 
\begin{equation}
    \Phi_s = \frac{N_s\mathcal{V}_{s}}{\vdomain} = \frac{N_s(\pi/6) D_s^3}{16000D_s^3} = 0.05,
\end{equation}
the settling velocity $U_s = 0.63$ and the Galilei number $\Galilei = 3.16$. The aspect ratio of the domain and the volume fraction of the spheres where chosen to match the laboratory experiments and the numerical simulations of \citet{Murch_Shaqfeh_2020}. The only varying parameter is the relaxation time of the polymers $\tau_\polymer$, which we set to $1.6$, $10$, $20$, $78.15$ and $155.72$, and its non-dimensional form, the Deborah number $\Deborah$ which takes the values $1.0$, $6.3$, $12.6$, $49.24$ and $98.49$, respectively (see table \ref{tab:results} in this article). 

\subsection{Numerical method}
The domain is discretised using second-order finite differences and a uniform Cartesian mesh of $256^2\times 512$ points. The momentum balance \eqref{eq:momentum_balance} is integrated with an Adams-Bashforth scheme and incompressibility is enforced by solving a Poisson equation for the pressure. The presence of the particles is accounted for through the Eulerian immersed boundary method (IBM) described by \citet{Hori_Takagi_Rosti_2022}. The same Adams-Bashforth scheme is used for the transport equation of the conformation tensor \eqref{eq:conformation_transport} \cite{Izbassarov_Rosti_Ardekani_Sarabian_Hormozi_Brandt_Tammisola_2018}, although the nonlinear fluxes of the upper convected derivative (e.g. the left-hand side of \eqref{eq:conformation_transport}) are reconstructed with a fifth-order weighted essentially non-oscillatory (WENO) scheme \cite{Sugiyama_Ii_Takeuchi_Takagi_Matsumoto_2011}. A matrix-logarithm formulation is employed to ensure the positive definiteness of the conformation tensor at high $\Deborah$ \cite{Fattal_Kupferman_2004}.  

The particle translational velocity is integrated at every time step to update the particle location $X_i$, where $U_i = \differential X_i/\differential t$. Inter-particle collisions are handled with the mass-spring-dash pot model \cite{Tsuji_Kawaguchi_Tanaka_1993,Costa_Boersma_Westerweel_Breugem_2015}. A repulsive force acts on the particles as soon as their volumes overlap, i.e. when the distance of their centres falls below $D_s$. 

The numerical algorithm is implemented in the in-house solver \href{https://www.oist.jp/research/research-units/cffu/fujin}{\textit{Fujin}} \cite{Rosti_2026}. The direct numerical simulations were performed using $256$ cores on the CPU-based cluster Deigo at the Okinawa Institute of Science and Technology.

\section{Results}

\begin{table}
\begin{ruledtabular}
\begin{tabular}{cccccccc}
& $\tau_\polymer$ & $\Deborah$ & $\averagespacetime{C_{kk}}$ & $\mathcal{L}/D_s$ & $\sigmavoro$ & $\displaystyle\frac{\averagespheretime{\mathcal{K}_\perp}}{U_s^2}$ & $\displaystyle\frac{\averagespheretime{\mathcal{K}_\parallel}}{U_s^2}$ \\
\tikz[baseline=-0.5ex]\draw[fill=black, draw=none] (-0.3em,-0.3em) rectangle (1.2em,0.6em); & $0$ & $0$ & n.a. & n.a. & $0.33$ & $0.0020$ & $0.0237$ \\
\tikz[baseline=-0.5ex]\draw[fill=De001, draw=none] (-0.3em,-0.3em) rectangle (1.2em,0.6em); & $1.6$ & $1$ & $3.04$ & $0.78$ & $0.40$ & $0.0035$ & $0.0348$ \\
\tikz[baseline=-0.5ex]\draw[fill=De006, draw=none] (-0.3em,-0.3em) rectangle (1.2em,0.6em); & $10$ & $6$ & $5.47$ & $2.35$ & $0.56$ & $0.0123$ & $0.1706$ \\
\tikz[baseline=-0.5ex]\draw[fill=De013, draw=none] (-0.3em,-0.3em) rectangle (1.2em,0.6em); & $20$ & $13$ & $31.30$ & $7.05$ & $1.18$ & $0.0254$ & $0.6560$ \\
\tikz[baseline=-0.5ex]\draw[fill=De050, draw=none] (-0.3em,-0.3em) rectangle (1.2em,0.6em); & $78.15$ & $50$ & $258.74$ & $9.37$ & $1.22$ & $0.0413$ & $1.6528$ \\
\tikz[baseline=-0.5ex]\draw[fill=De100, draw=none] (-0.3em,-0.3em) rectangle (1.2em,0.6em); & $155.72$ & $100$ & $685.63$ & $8.91$ & $1.11$ & $0.0451$ & $2.5152$ \\
\end{tabular}
\end{ruledtabular}
\caption{Results of the direct numerical simulations for the values of the polymer relaxation time $\tau_\polymer$ and Deborah number $\Deborah$ considered in this work. The colors describing each case are shown in the leftmost column. The remaining columns list the space-time average of $C_{kk}$, its correlation length of $\mathcal{L}$, the standard deviation of the Voronoi volumes $\sigmavoro$, the kinetic energy of the spheres in the horizontal direction $\mathcal{K}_\perp$ and the kinetic energy of the spheres in the vertical direction $\mathcal{K}_\parallel$.\label{tab:results}}
\end{table}

Table \ref{tab:results} reports the numerical values used to prepare the figures in the manuscript.

\subsection{Voronoi tessellation}
\begin{figure}
    \centering
    \includegraphics[width=\linewidth]{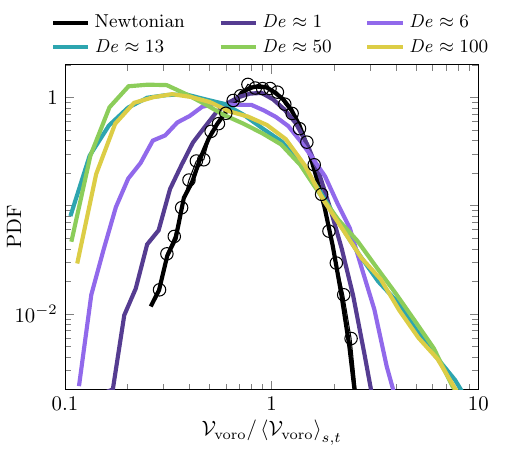}
    \caption{Probability density function (PDFs) of the normalized Voronoi volumes $\vvoro/\averagespheretime{\vvoro}$ for the Newtonian (black) and viscoelastic (colored) cases $\Deborah$ listed in table \ref{tab:results}. The PDF obtained for $N_s$ spheres which were distributed through a random Poisson process (RPP) is plotted with the hollow circles \protect\hollowlinesymbol\ for comparison.}
    \label{fig:voronoi}
\end{figure}

We quantified particle clustering by computing the Voronoi tessellation of the sphere-laden domain. The latter is partitioned in a set of subvolumes, one for each sphere, which are referred to as Voronoi volumes. A sphere's Voronoi volume $\vvoro$ contains all the points that are closer to the that it to any other. Thus, small and large Voronoi volumes are found in regions of high- and low-concentration, respectively. By construction, the ensemble average of the Voronoi volumes is constant in time and equal to the inverse of the mean concentration $\averagespheretime{\vvoro}=\mathcal{\vdomain}/N_s$. A Voronoi tessellation is therefore local in space and time and its structure depends only on the instantaneous location of the spheres $X_i(t)$ \citep{Monchaux_Bourgoin_Cartellier_2010,Monchaux_2012,Fiabane_Zimmermann_Volk_Pinton_Bourgoin_2012}. We computed the tessellation with an in-house code based on the \href{https://math.lbl.gov/voro++/}{Voro++} library, which handles three-dimensional domains with triple periodicity \cite{Lu_Lazar_Rycroft_2023}, and repeated the procedure for every time instance. The probability density functions (PDFs) of the normalized Voronoi volumes $\vvoro/\averagespheretime{\vvoro}$ are shown in figure \ref{fig:voronoi} for the Newtonian (black) and viscoelastic (color) cases. These PDFs are compared with a benchmark PDF (gray) obtained from a set of $N_s$ spheres whose position $X_i$ is assigned through a random Poisson process (RPP) within the same domain $\left(20D_s\right)^2\times40D_s$. Because the spheres generated through a RPP are not concentrated, the departure of the colored PDFs in figure \ref{fig:voronoi} from the RPP and towards smaller and larger volumes indicates the presence of clusters and voids \cite{Monchaux_Bourgoin_Cartellier_2010}. While the PDF of the Newtonian case (black) collapses perfecly on the RPP one (hollow circles), the PDFs of the viscoelastic case broaden for $\Deborah$ up to $13$ and include volumes that are much smaller or much larger than those of the RPP distribution. The inset in figure \ref{fig:voronoi} shows the normalized standard deviation $\sigmavoro/\averagespheretime{\vvoro}$ which quantifies the degree of clustering \cite{Monchaux_Bourgoin_Cartellier_2010}. It increases monotonically for $\Deborah\lesssim13$, saturates for $\Deborah\geq13$ and decreases slightly at $\Deborah\approx100$. 

\subsection{Scale-by-scale energy budget and kinetic energy spectra}

\begin{figure*}
    \centering
    \includegraphics[width=\linewidth]{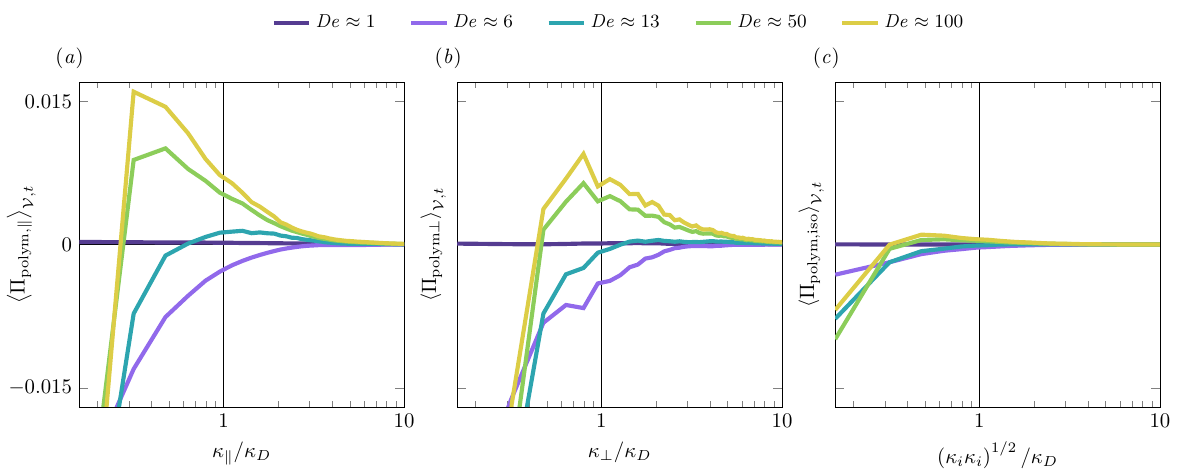}
    \caption{Space-time average of the work of the polymer stresses for increasing Deborah number $\Deborah$. The panels show the vertical contribution $\Pi_{\polymer\parallel}$ \eqref{eq:polymer_horizontal} as a function of the vertical wavenumber $\kappa_\parallel$ (\textit{a}), the horizontal contribution $\Pi_{\polymer\perp}$ \eqref{eq:polymer_horizontal} as a function of the horizontal wavenumber $\kappa_\perp$ (\textit{b}) and the isotropic integral over a spherical shell $\Pi_{\polymer,\text{iso}}$ as a function of the radial wavenumber $(\kappa_i\kappa_i)^{1/2}$ (\textit{c}). The dashed and dash-dot lines represent the $-3$ and $-4$ scaling, respectively.}
    \label{fig:polymer_stresses_complete}
\end{figure*}
\begin{figure*}
    \centering
    \includegraphics[width=\linewidth]{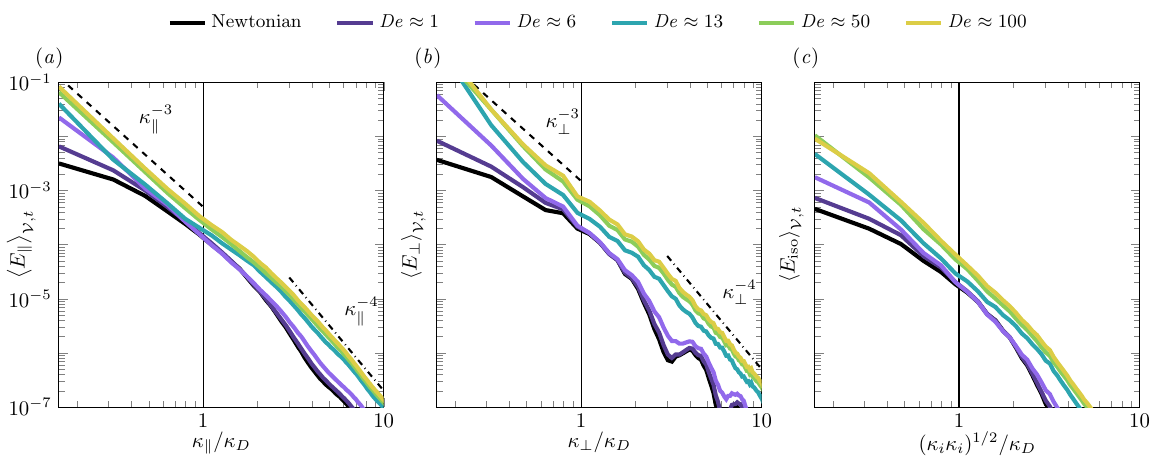}
    \caption{Space-time average of the turbulent kinetic energy spectra of the fluid for increasing Deborah number $\Deborah$. The panels show the spectra of the vertical velocity $E_\parallel$ \eqref{eq:spectra_horizontal} as a function of the vertical wavenumber $\kappa_\parallel$ (\textit{a}), the spectra of the horizontal velocity $E_\perp$ \eqref{eq:spectra_horizontal} as a function of the horizontal wavenumber $\kappa_\perp$ (\textit{b}) and the isotropic spectra $E_{\text{iso}}$ integrated over a spherical shell as a function of the radial wavenumber $(\kappa_i\kappa_i)^{1/2}$ (\textit{c}). The dashed and dash-dot lines represent the $-3$ and $-4$ scaling, respectively.}
    \label{fig:energy_spectra_complete}
\end{figure*}

The terms of the scale-by-scale energy budget are obtained by contracting the Fourier transform of the momentum balance equation \eqref{eq:momentum_balance} with that of the fluid velocity $\hat u_i$. The result is the equation of the kinetic energy in the spectral form which, in dimensional form, reads
\begin{multline}
    \p_t\Phi_{uu} = \pi\left(\kappa_i\right) + \pi_{\particle}\left(\kappa_i\right) + \pi_{\gravity}\left(\kappa_i\right) + \\ 
    + \pi_{\pressure}\left(\kappa_i\right) + \varepsilon_{uu}\left(\kappa_i\right) + \pi_{\polymer}\left(\kappa_i\right),
\end{multline}
where the superscript $\ast$ denotes the complex conjugate and summation over repeated indices is implied. Here, $\pi$ denotes the energy flux of the nonlinear terms, $\pi_\particle$ the work of the particle forcing, $\pi_\gravity$ is the work of the gravity and buoyancy forces, $\pi_\pressure$ the work of the pressure gradient, $\mathcal{D}_\viscous$ the viscous dissipation and $\pi_\polymer$ the work of the polymer stresses:
\begin{subequations}\label{eq:scale_by_scale_terms}
\begin{equation}
    \pi \left(\kappa_k\right) = \kappa_k \left( \Re \widehat{u_ku_i} \Im \hat u_i - \Im \widehat{u_ku_i} \Re \hat u_i\right),
\end{equation}
\begin{equation}
    \pi_\particle \left(\kappa_k\right) = \Re\hat f_k^\particle\Re\hat u_k + \Im\hat f_k^\particle\Im\hat u_k,
\end{equation}
\begin{equation}
    \pi_\gravity \left(\kappa_k\right) = \Re\hat f_k^\gravity\Re\hat u_k + \Im\hat f_k^\gravity\Im\hat u_k,
\end{equation}
\begin{equation}
    \pi_\pressure\left(\kappa_k\right) = \frac{\kappa_k}{\rho_f}\left(\Re\hat u_k\Im\hat p - \Re\hat p\Im\hat u_k\right),
\end{equation}
\begin{equation}
    \mathcal{D}_\viscous\left(\kappa_k\right) = - \nu_f \kappa^2 \left(\Re \hat u_k\Re \hat u_k + \Im \hat u_k\Im \hat u_k\right),
\end{equation}
\begin{multline}\label{eq:scale_by_scale_terms_polymer}
    \pi_\polymer\left(\kappa_k\right) = \frac{\kappa_k\mu_\polymer}{\tau_\polymer\rho_f} \left( \Re\hat C_{ik} \Im \hat u_i + \right.\\
    \left.- \Im \hat C_{ik} \Re \hat u_i\right).
\end{multline}
\end{subequations}
where $\Re$ and $\Im$ denote the real and imaginary parts, respectively. The kinetic energy spectrum and the viscous dissipation are computed from the velocity product $\hat u_i\hat u_i^\ast$
\begin{equation}
    \Phi_{uu}\left(\kappa_k\right) = \Re \hat u_k\Re \hat u_k + \Im \hat u_k\Im \hat u_k.
\end{equation}
By virtue of the axisymmetry of the system one can decompose the terms \eqref{eq:scale_by_scale_terms} to account for the horizontal ($\perp$, $k=1,2$) and vertical ($\parallel$, $k=3$) contributions. For instance, the horizontal and vertical contributions for the polymer term \eqref{eq:scale_by_scale_terms_polymer} expand as
\begin{widetext}
\begin{subequations}
    \begin{multline}
        \pi_{\polymer\perp}\left(\kappa_x,\kappa_y,\kappa_z\right) = \\ = \frac{\kappa_x\mu_\polymer}{\tau_\polymer\rho_f} \left( \Re\hat C_{xx} \Im \hat u_x - \Im \hat C_{xx} \Re \hat u_x + \Re\hat C_{yx} \Im \hat u_y - \Im \hat C_{yx} \Re \hat u_y + \Re\hat C_{zx} \Im \hat u_z - \Im \hat C_{zx} \Re \hat u_z\right) + \\
        + \frac{\kappa_y\mu_\polymer}{\tau_\polymer\rho_f} \left( \Re\hat C_{xy} \Im \hat u_x - \Im \hat C_{xy} \Re \hat u_x + \Re\hat C_{yy} \Im \hat u_y - \Im \hat C_{yy} \Re \hat u_y + \Re\hat C_{zy} \Im \hat u_z - \Im \hat C_{zy} \Re \hat u_z\right) ,
    \end{multline}
    \begin{multline}
        \pi_{\polymer\parallel}\left(\kappa_x,\kappa_y,\kappa_z\right) = \\ 
        = \frac{\kappa_z\mu_\polymer}{\tau_\polymer\rho_f} \left( \Re\hat C_{xz} \Im \hat u_x - \Im \hat C_{xz} \Re \hat u_x + \Re\hat C_{yx} \Im \hat u_y \right. - \Im \hat C_{yz} \Re \hat u_y + \left.\Re\hat C_{zx} \Im \hat u_z - \Im \hat C_{zz} \Re \hat u_z\right).
    \end{multline}
\end{subequations}
\end{widetext}
The horizontal and vertical contributions of the spectrum are
\begin{subequations}
    \begin{multline}
        \Phi_{uu\perp}\left(\kappa_x,\kappa_y,\kappa_z\right) = \Re \hat u_x\Re \hat u_x + \Im \hat u_x\Im \hat u_x + \\
        + \Re \hat u_y\Re \hat u_y + \Im \hat u_y\Im \hat u_y,
    \end{multline}
    \begin{equation}
        \Phi_{uu\parallel}\left(\kappa_x,\kappa_y,\kappa_z\right) = \Re \hat u_z\Re \hat u_z + \Im \hat u_z\Im \hat u_z,
    \end{equation}
\end{subequations}
and so on. The resulting terms are averaged over horizontal planes or along the vertical axis by invoking the shifting property of the Dirac delta \citep[\S 6.5]{Pope_2000}
\begin{widetext}
\begin{subequations}
\begin{equation}\label{eq:polymer_parallel}
    \Pi_{\polymer\perp}\left(\kappa_\perp\right) = \int_{-\infty}^{+\infty}\int_0^{+\infty} \left[\pi_{\polymer\perp}\left(\kappa_\alpha,\kappa_\parallel\right) \delta\left(\kappa_\alpha - \kappa_{\perp}\right) \differential\kappa_\alpha\right] \differential\kappa_\parallel,
\end{equation}
\begin{equation}\label{eq:spectra_parallel}
    E_{\perp}\left(\kappa_\perp\right) = \int_{-\infty}^\infty \int_0^\infty \left[\Phi_{uu}\left(\kappa_\alpha,\kappa_\parallel\right) \delta\left(\kappa_\alpha - \kappa_{\perp}\right) \differential\kappa_\alpha\right] \differential\kappa_\parallel,
\end{equation}
\begin{equation}\label{eq:polymer_horizontal}
    \Pi_{\polymer\parallel}\left(\kappa_\parallel\right) = \int_0^{+\infty}\int_0^{+\infty} \left[\pi_{\polymer\parallel}\left(\kappa_\alpha,\kappa_\parallel\right) \delta\left(\kappa_\alpha - \kappa_{\perp}\right) \differential\kappa_\alpha\right] \differential\kappa_\perp,
\end{equation}
\begin{equation}\label{eq:spectra_horizontal}
    E_{\parallel}\left(\kappa_\parallel\right) = \int_0^{+\infty}\int_0^{+\infty} \left[\Phi_{uu}\left(\kappa_\alpha,\kappa_\parallel\right) \delta\left(\kappa_\alpha - \kappa_{\perp}\right) \differential\kappa_\alpha\right] \differential\kappa_\perp.
\end{equation}
\end{subequations}
\end{widetext}
where we used the horizontal wavenumber $\kappa_\perp = (\kappa_x^2 + \kappa_y^2)^{1/2} = \left(\kappa_\alpha\kappa_\alpha\right)^{1/2}$ and the vertical wavenumber $\kappa_\parallel=\kappa_z$. The space-time averages of the polymer stress term and the kinetic energy spectra are drawn with the solid curves in figure \ref{fig:polymer_stresses_complete} and \ref{fig:energy_spectra_complete}, respectively. The plots of the vertical contribution (\ref{fig:polymer_stresses_complete}\textit{a}) and horizontal contribution (\ref{fig:polymer_stresses_complete}\textit{b}) are compared to the isotropic integral over a spherical shell of radius $\kappa=\left(\kappa_i\kappa_i\right)^{1/2}$. The vertical and horizontal contribution show the same qualitative trends, albeit slight differing in magnitude. Notably, the low-wavenumber scaling $-3$ (dashed) and the high-wavenumber scaling $-4$ (dash dot) of the spectra in figure \ref{fig:energy_spectra_complete} are only present in the horizontal and vertical contributions and not in the isotropic spectra (\ref{fig:energy_spectra_complete}\textit{c}).

\subsection{Structure functions}
\begin{table*}
\begin{ruledtabular}
\begin{tabular}{cccccccc}
& $\Deborah$ & $n=1$ & $n=2$ & $n=3$ & $n=4$ & $n=5$ & $n=6$ \\
\tikz[baseline=-0.5ex]\draw[fill=black, draw=none] (-0.3em,-0.3em) rectangle (1.2em,0.6em); & $0$ & $1.7636\pm0.0043$ & $3.4715\pm0.0094$ & $5.1249\pm0.0141$ & $6.7376\pm 0.0171$ & $8.3264\pm 0.0181$ & $9.9040\pm 0.0180$ \\
\tikz[baseline=-0.5ex]\draw[fill=De001, draw=none] (-0.3em,-0.3em) rectangle (1.2em,0.6em); & $1$ & $1.7613\pm0.0023$ & $3.4635\pm0.0060$ & $5.1080\pm0.0110$ & $6.7086\pm0.0177$ & $8.2826\pm0.0263$ & $9.8435\pm0.0370$ \\
\tikz[baseline=-0.5ex]\draw[fill=De006, draw=none] (-0.3em,-0.3em) rectangle (1.2em,0.6em); & $6$ & $1.7233\pm0.0040$ & $3.3404\pm0.0102$ & $4.8523\pm0.0211$ & $6.2676\pm0.0427$ & $7.5978\pm0.0852$ & $8.8544\pm0.1737$ \\
\tikz[baseline=-0.5ex]\draw[fill=De013, draw=none] (-0.3em,-0.3em) rectangle (1.2em,0.6em); & $13$ & $1.6007\pm0.0256$ & $2.9995\pm0.0574$ & $4.2033\pm0.0934$ & $5.1890\pm0.1513$ & $5.9918\pm 0.2252$ & $6.6794\pm0.3000$ \\
\tikz[baseline=-0.5ex]\draw[fill=De050, draw=none] (-0.3em,-0.3em) rectangle (1.2em,0.6em); & $50$ & $1.6106\pm0.0251$ & $2.9824\pm0.0565$ & $4.1279\pm0.1056$ & $4.9733\pm0.3069$ & $5.6041\pm0.5135$ & $6.1513\pm0.6559$ \\
\tikz[baseline=-0.5ex]\draw[fill=De100, draw=none] (-0.3em,-0.3em) rectangle (1.2em,0.6em); & $100$ & $1.5964\pm0.0261$ & $2.9828\pm0.0748$ & $4.1170\pm0.1517$ & $4.8192\pm0.2357$ & $5.1638\pm0.4847$ & $5.4816\pm0.7135$ \\
\end{tabular}
\end{ruledtabular}
\caption{Temporal average and standard deviation of the exponents $\zeta_n$ of the structure function of the second differences $\Sigma_n=O(r_\parallel^{\zeta_n})$ for small separations $r_\parallel$.\label{tab:exponents}}
\end{table*}

If a homogeneous and statistically steady-state velocity field $u(x)$ is differentiable $p$ times, its energy spectrum $E\left(\kappa\right)$ satisfies the relation \cite[G.11]{Pope_2000}
\begin{equation}\label{eq:spectrum_differentiability_integral}
    \average{\left(\frac{\differential^p u}{\differential x^p}\right)^2} = \int_{0}^{+\infty} \kappa^{2n} E\left(\kappa\right) \differential\kappa.
\end{equation}
Let us assume that the spectrum is a power law of the form $E\left(\kappa\right) = \mathcal{E}\kappa^{-\alpha}$. The integral on the right-hand side of \eqref{eq:spectrum_differentiability_integral} converges in the high-wavenumber limit for $\alpha>1$ if $u$ is rough ($p=0$) and for $\alpha>3$ if it is one-time differentiable ($p=1$). The velocity expands as \cite{Biferale_Cencini_Lanotte_Vergni_Vulpiani_2001}
\begin{equation}\label{eq:one-time-differentiable-taylor-expansion}
    u\left(x+r\right) - u\left(x\right) \sim \frac{\p u}{\p r} r + a\left(x\right) r^h,
\end{equation}
where $h$ is the nontrivial exponent \cite{Singh_Perlekar_Mitra_Rosti_2024}. The behavior of the one-time differentiable field is distilled in the longitudinal structure functions of the second differences of order $n$ \cite{Biferale_Cencini_Lanotte_Vergni_2003}
\begin{equation}\label{eq:structure_function_second_difference_nontrivial_exponent}
    \Sigma_n = \averagespace{\left(\delta_u^2\right)^n} = O\left( r^{\zeta_n}\right),
\end{equation}
where $\delta_u^2= u\left(x+r\right) - 2u\left(x\right) + u\left(x-r\right)$ and $\zeta_p=ph$. The second-order structure function of the second differences $\Sigma_2$ can be expressed in terms of the velocity autocorrelation $\averagespace{u\left(x\right)u\left(x+r\right)}$ and Fourier transformed \cite[G.4]{Pope_2000}
\begin{equation}\label{eq:second_order_structure_function_second_difference}
    \Sigma_2 = 4\mathcal{E} r^{\alpha-1} \int_0^{+\infty} \left(\kappa r\right)^{-\alpha} \left[1 - \cos\left(\kappa r\right)\right]^2 \differential\left(\kappa r\right).
\end{equation}
For small $\kappa r\ll1$, the integral in \eqref{eq:second_order_structure_function_second_difference} converges if $\alpha<5$. Equating the powers in \eqref{eq:structure_function_second_difference_nontrivial_exponent} and \eqref{eq:second_order_structure_function_second_difference} yields $\alpha = 2h + 1$ \cite{Biferale_Cencini_Lanotte_Vergni_2003} and $\zeta_n = n\left(\alpha-1\right)/2$. For $3<\alpha<5$, the second-order moment of \eqref{eq:one-time-differentiable-taylor-expansion} takes the form $S_2 = \averagespace{\left[u\left(x+r\right) - u\left(x\right)\right]^2}\sim A r^2 + Br^{\alpha-1}$, where $A$ and $B$ are real constants \cite{Bernard_2000,Singh_Perlekar_Mitra_Rosti_2024}. In these cases, the term $Br^{\alpha-1}$ is subdominant at small $r$ and $S_2=O(r^2)$ for any $\alpha$ and the intermittent behavior of the velocity field is distilled in the structure function \eqref{eq:structure_function_second_difference_nontrivial_exponent} $\Sigma_2 \sim Cr^{\alpha-1}$. 

To compute the exponents $\zeta_n$, we followed the procedure outlined by \citet{Singh_Perlekar_Mitra_Rosti_2024}. We first computed the structure functions for every available velocity dataset using the definition $\Sigma_n(r_\parallel,t)=\averagespace{[\Delta_\parallel^2(r_\parallel,t)]^n}=O\left(r_\parallel^{\zeta_n}\right)$, where $\Delta_\parallel^2(r_\parallel,t) = u_\parallel(\boldsymbol{x} + r_\parallel\boldsymbol{e}_\parallel,t) - 2u_\parallel(\boldsymbol{x},t) + u_\parallel(\boldsymbol{x}-r_\parallel\boldsymbol{e}_\parallel,t)$ \citep{Biferale_Cencini_Lanotte_Vergni_2003}. We then identified a suitable scaling range $0.1562\leq r_\parallel/D_s<0.7812$ at small $r_\parallel$ to extract the exponents by using the relation \citep{Singh_Perlekar_Mitra_Rosti_2024}
\begin{equation}\label{eq:extracting_exponents}
    \zeta_n(t) = \frac{\differential \ln\Sigma_n}{\differential \ln r_\parallel}.
\end{equation}
The temporal average $\averagetime{\zeta_n(t)}$ and standard deviation $\left(\average{\zeta_n(t)^2}-\averagetime{\zeta_n(t)}^2\right)^{1/2}$ are listed in table \ref{tab:exponents} for $n\leq6$.

\bibliography{biblio.bib}